\documentclass[12pt,letterpaper]{article}

\usepackage{amsmath,amssymb,array,calc,rotating,epsfig,psfrag, amscd}
\usepackage{color}
\usepackage[
      colorlinks=true,
      urlcolor=blue,    
      filecolor=blue,     
      citecolor=red,
      pdfstartview=FitV,
       bookmarksopen=true    
      ]{hyperref}
\usepackage[left=2cm,top=1cm,right=3cm,nohead]{geometry}

\newcommand{\Blp}{{\Big (}}
\newcommand{\Brp}{{\Big )}}

\newcommand{\slb}{{\rm [}}
\newcommand{\srb}{{\rm ]}}

\def\lam{\lambda}

\newcommand{\bea}{\begin{eqnarray}}
\newcommand{\eea}{\end{eqnarray}}
\newcommand{\be}{\begin{equation}}
\newcommand{\ee}{\end{equation}}
\newcommand{\barr}{\begin{array}}
\newcommand{\earr}{\end{array}}

\newcommand{\ZZ}{{\mathbb Z}}

\newcommand{\half}{\frac{1}{2}}

\newcommand{\del}{\partial}

\newcommand{\tphi}{\tilde{\phi}}

\newcommand{\txi}{\tilde{\xi}}

\newcommand{\csch}{{\rm csch}}

\newcommand{\non}{\nonumber}

\newcommand{\bpm}{\begin{pmatrix}}
\newcommand{\epm}{\end{pmatrix}}
 \newcommand{\bitem}{\begin{itemize}}
 \newcommand{\eitem}{\end{itemize}}

\definecolor{cardinal}{rgb}{0.6,0,0}
\definecolor{darkgreen}{rgb}{0,0.5,0}
\definecolor{golden}{rgb}{0.92, 0.7, 0}
\definecolor{midnight}{rgb}{0, 0, 0.5}
\definecolor{darkblue}{rgb}{0.2, 0, 0.8}

\newcommand{\beq}{\begin{equation}\begin{aligned}}
\newcommand{\eeq}{\end{aligned}\end{equation}}
\newcommand{\nn}{\nonumber}

\newcommand{\fc}{\phi^c}
\newcommand{\fd}{\phi^d}

\newcommand{\fo}{\phi_0}

\newcommand{\fua}{\phi_1^a}
\newcommand{\fub}{\phi_1^b}

\begin{document}

\begin{flushright}
IPhT-t11/019
\end{flushright}
\begin{center}
\end{center}

\vspace{0.5cm}
\begin{center}
\baselineskip=13pt {\LARGE \bf{On Metastable Vacua and the \\ Warped Deformed Conifold: \\ Analytic Results}\\}
 \vskip1.5cm 
 Iosif Bena$^{*}$, Gregory Giecold$^{*}$, Mariana Gra\~na$^{*}$, Nick Halmagyi$^{* \dagger}$ and Stefano Massai$^{*}$\\ 
 \vskip0.5cm
$^{*}$\textit{Institut de Physique Th\'eorique,\\
CEA Saclay, CNRS URA 2306,\\
F-91191 Gif-sur-Yvette, France}\\
\vskip0.8cm
$^{\dagger}$\textit{Laboratoire de Physique Th\'eorique et Hautes Energies,\\
Universit\'e Pierre et Marie Curie, CNRS UMR 7589, \\
F-75252 Paris Cedex 05, France}\\
\vskip0.5cm
iosif.bena, gregory.giecold, mariana.grana, stefano.massai@cea.fr\\
\vskip0.2cm
halmagyi@lpthe.jussieu.fr \\ 
\end{center}
\vskip1.5cm
\begin{abstract}

Continuing the programme of constructing the backreacted solution corresponding to smeared anti--D3 branes in the warped deformed conifold, we solve analytically the equations governing the space of first--order deformations around this solution. We express the results in terms of at most three nested integrals. These are the simplest expressions for the space of $SU(2) \times SU(2) \times \ZZ_2$--invariant deformations, in which the putative solution for smeared anti--D3 branes must live. We also explain why one cannot claim to identify this solution without fully relating the coefficients of the infrared and ultraviolet expansions of the deformation modes. The analytic solution we find is the first step in this direction.

  \end{abstract}

\newpage

\section{Introduction and Motivation}

Anti--D3 branes in Klebanov-Strassler (KS) warped deformed conifold throats~\cite{Klebanov:2000hb} are a key ingredient in
string model building and string cosmology, where they are used both for lifting AdS to de Sitter solutions~\cite{Kachru:2003aw}, and to construct models of inflation using D3 branes moving in KS--like geometries~\cite{Kachru:2003sx}. 

In a previous paper~\cite{Bena:2009xk}, three of the authors solved for the space of $SU(2) \times SU(2) \times \ZZ_2$--invariant deformations, in which the would--be solution describing anti--D3 branes smeared at the tip of the warped deformed conifold must live. The equations parameterizing this space of deformations were found in~\cite{Borokhov:2002fm}, and by repackaging them and finding their homogeneous solutions, one can write down the implicit solution to the full space of deformations in terms of nested integrals~\cite{Bena:2009xk}.
Performing these nested integrals is no easy task. To analyze the physics of the backreacted solution sourced by antibranes and determine whether this solution is dual to a metastable vacuum of the dual gauge theory (as conjectured by~\cite{Kachru:2002gs} following a probe analysis), the authors of~\cite{Bena:2009xk} found the explicit infrared and ultraviolet expansions of all the modes\footnote{The UV expansion of a subset of the invariant modes for the Klebanov--Tseytlin solution has been considered in~\cite{DeWolfe:2008zy} and the IR expansion of a deformation around KS has been looked at in~\cite{McGuirk:2009xx}.}. This analysis revealed that the candidate anti--D3 brane solution must have a certain infrared singularity. If this singularity is physical, then the space of solutions implicitly--solved--for in~\cite{Bena:2009xk} captures the first--order backreaction of antibranes in KS. If this singularity is pathological, the analysis of~\cite{Bena:2009xk} implies that the backreaction of anti--D3 branes in KS cannot be treated in perturbation theory\footnote{A similar result was obtained when investigating the backreaction of putative brane--engineered metastable vacua~\cite{Bena:2006rg}, or the backreaction of anti--M2 branes in a warped Stenzel background~\cite{Bena:2010gs}.}. 

As discussed in detail in~\cite{Bena:2009xk} there are many arguments why this singularity may be physical or not, but so far there exists no criterion for accepting it that does not force one to also accept clearly unphysical singularities like that of the negative--mass Schwarzschild solution\footnote{An example of such a criterion is accepting so-called ``normalizable,'' singularities whose energy density diverges, but whose energy is finite.}. Solving the fully--backreacted solution for these anti--D3 branes could settle the issue, but this involves solving eight coupled second--order nonlinear differential equations. A more indirect way to determine whether this singularity is physical would be to allow it, and then explore whether the physics of the resulting solution is consistent with the physics one expects from anti--D3 branes. To do this, one must set all the other unphysical IR singularities and all the UV non--normalizable modes to zero (which was done in~\cite{Bena:2009xk}), and then relate the various coefficients of the infrared and ultraviolet expansions. This will fix all the parameters of the solution in terms of one parameter, the number of anti--D3 branes, and yield the relation between the charge of the would--be antibranes and the force they exert on a probe D3 brane. Since this force is known from KKLMMT~\cite{Kachru:2003sx} and is not screened~\cite{Bena:2010ze}, if the force--to--charge ratio is the correct one for anti--D3 branes, this indicates that the singularity is most likely physical, and the solution describes indeed the backreaction of anti--D3 branes. If this ratio is not the correct one then the solution has nothing to do with antibranes, and is most likely unphysical. Note that obtaining this ratio from the solution is a nontrivial process, which involves matching all the infrared and ultraviolet integration constants in the expansions of the modes.

One way to obtain this ratio would be to perform numerically all the nested integrals that appear in the solution of~\cite{Bena:2009xk} and relate the corresponding ultraviolet and infrared integration constants. However, this gets complicated, in particular because most of the integrals either diverge or decay very fast. What simplifies the analysis is the observation that one of the modes $(\xi_1)$ is proportional to the warp factor in the Klebanov--Strassler solution. This simplifies the expressions of the integrals appearing in the functions 
$\xi_a$, which in the Borokhov--Gubser formalism are the ``conjugate momenta'' to the physical perturbation fields $\phi_a$~\cite{Borokhov:2002fm}. The second key observation is that most of the nested double integrals that appear at various stages of solving the Borokhov--Gubser equations can be integrated by parts, and hence can be expressed in terms of single integrals. 

In this note we show that by making judicious use of these two observations, one can express all eight conjugate momentum functions $\xi_a$ as linear combinations of simple integrals. Furthermore, with the exception of the perturbation to the warp factor, all the other seven perturbation modes can be expressed in terms of just two nested integrals. The expression for the perturbed warp factor is the most complicated of them all, and can be expressed in terms of three nested integrals. This is a huge improvement from the implicit solution of~\cite{Bena:2009xk}, where the most complicated of the $\xi_a$ was expressed in terms of four nested integrals, and the warp factor ($\phi_4$) was expressed in terms of eight of them.

Hence, the expressions presented in this paper correspond to the simplest solution to the space of $SU(2) \times SU(2) \times \ZZ_2$--invariant deformations of the Klebanov--Strassler background, and should include both the putative antibrane solution, as well as other perturbations of the Klebanov--Strassler field theory with various non-normalizable modes that have no R--charge (see for example
~\cite{Kuperstein:2003yt,Berg:2005pd,Berg:2006xy}). 

In an upcoming paper we will use these results to perform the numerical integration of the putative anti--D3 brane solution of~\cite{Bena:2009xk}, allowing for the extra infrared singularity. This will relate the ultraviolet and the infrared integration constants and allow us to determine whether the ratio between the anti--D3 charge of the solution and the force on a probe D3 brane is the correct one. 

\vspace{.2cm}

\noindent {\bf Note:} While we were preparing this paper for submission we became aware of~\cite{Dymarsky:2011pm}, in which some of the nested integrals obtained in~\cite{Bena:2009xk} were explicitly performed. In the solution of~\cite{Bena:2009xk} the most complicated field can be expressed in terms of 8 nested integrals, in \cite{Dymarsky:2011pm} this is expressed in terms of 7 nested integrals, and in the present paper in terms of 3. Furthermore, as it is clear from the discussion above, we believe that it is premature to claim that a solution describes anti--D3 branes without first computing the charge--to--force ratio and making sure it is the correct one. 

\newpage

\section{Analytic Solution for Deformations of the Warped Deformed Conifold}

In this section we present our analytic formulae for the full set of   $SU(2) \times SU(2) \times \ZZ_2$--invariant 
deformation space around the Klebanov--Strassler solution\footnote{See~\cite{Bena:2010pr,Cassani:2010na} for a rigorous derivation of a consistent truncation on $T^{1,1}$ which includes several additional modes.}. The method follows that of~\cite{Bena:2009xk}, but here we present numerous analytical improvements.

We use the ansatz for the supergravity background fields proposed by Papadopoulos and Tseytlin (PT)~\cite{Papadopoulos:2000gj}, which is the most general ansatz consistent with the $SU(2)\times SU(2)\times \ZZ_2$--symmetry of the Klebanov--Strassler background:
\beq\label{PTmetric}
ds_{10}^2 = e^{2\, A+2\, p-x}\, ds_{1,3}^2 + e^{-6\, p-x}\, d\tau^2 + e^{x+y}\, \left( g_1^2 + g_2^2 \right) + e^{x-y}\, \left( g_3^2 + g_4^2 \right) + e^{-6\, p-x}\, g_5^2 \ ,
\eeq
where all the functions depend on the variable $\tau$.  The fluxes and dilaton are
\bea\label{PTfluxes}
 H_3 &=& \tfrac12 \, \left( k - f \right)\, g_5 \wedge \left( g_1 \wedge g_3+ g_2 \wedge g_4 \right) + d\tau \wedge \left( f'\, g_1 \wedge g_2+ k'\, g_3 \wedge g_4 \right) \ , \nn \\
 F_3&=& F\, g_1 \wedge g_2 \wedge g_5 + \left( 2\, P - F \right) \, g_3\wedge g_4 \wedge g_5 +
 F' \, d\tau \wedge \left( g_1 \wedge g_3 + g_2 \wedge g_4 \right) \ , \\
 F_5 &= & {\cal F}_5 + * {\cal F}_5 \, , \qquad {\cal F}_5 = \left[ k\, F + f\, \left( 2\, P -F \right) \right] \, g_1 \wedge g_2 \wedge g_3 \wedge g_4 \wedge g_5  \ ,\nn \\
 \Phi&=& \Phi(\tau)\, , \qquad C_0 = 0 \, ,
 \eea
where $P$ is a constant while $f,k$ and $F$ are functions of $
\tau$ and a prime denotes a derivative with respect to $\tau$.

\subsection{The Borokhov--Gubser method and the zeroth--order background}

The method introduced in~\cite{Borokhov:2002fm} relies on the existence of a superpotential $W$ whose square gives the potential, namely
 \be
 V(\phi) = \frac{1}{8}\, G^{ab}\, \frac{\del W}{ \del \phi^a}\, \frac{\del W}{ \del \phi^b} \, .
 \ee
 The fields $\phi^a$ ($a=1,...,n$) are expanded around their supersymmetric background value in the form 
\beq
 \label{split}
 \phi^a = \phi^a_0 + \phi^a_1(X) + {\cal O}(X^2)\, ,
 \eeq
where $X$ represents the set of perturbation parameters, $\phi^a_1$ is linear in them, and $\phi^a_0$ are the functions in the Klebanov--Strassler solution, written explicitly below.
The method amounts to splitting $n$ second--order equations ($n=8$ for us), into $2n$ first--order ones, out of which $n$ of them (those for the conjugate momenta $\xi_a$) form a closed set. The conjugate momenta $\xi_a$ are defined as
\beq
\label{xidef}
\xi_a \equiv G_{ab}(\phi_0)\, \left( \frac{d\phi_1^b}{d\tau} - M^b_{\ d}(\phi_0)\, \phi_1^d \right) \ , \qquad M^b{}_d\equiv\frac12 \, \frac{\partial}{\partial \fd}\, \left( G^{bc}\, \frac{\partial W}{\partial \fc} \right) \, .
\eeq
They are linear in the expansion parameters $X$ and non--zero for non--supersymmetric solutions only. The set $(\xi_a,\phi^a)$
satisfies the equations:
\bea
\frac{d\xi_a}{d\tau} + \xi_b\, M^b{}_a(\fo) &=& 0 \, ,  \label{xieq} \\
\frac{d\fua}{d\tau} - M^a{}_b(\fo)\, \fub &=& G^{ab}\, \xi_b \label{phieq} \, .
\eea
Note that equations~\eqref{phieq} are just a rephrasing of the definition of the $\xi_a$ in~\eqref{xidef}, while the equations in~\eqref{xieq}
imply the equations of motion~\cite{Borokhov:2002fm}. The functions $\xi_{a}$ should additionally satisfy the zero--energy condition
\beq \label{ZEC}
\xi_a\, \frac{d\fo^a}{d \tau} = 0 \, .
\eeq

We will denote the set of functions $\phi^a$, $a=1,...,8$ of the PT ansatz in the following order
\be
\label{phidef}
 \phi^a=(x,y,p,A,f,k,F,\Phi) \, .
 \ee
The field--space metric in~\eqref{xidef} is 
\bea
\label{fieldmetric}
G_{ab}\, \phi^{\prime a}\, \phi^{\prime b} &=& e^{4\, p + 4\, A}\, \Big[ x'^2 + \frac12 \, y'^2 + 6\, p'^2 - 6\, A'^2 + \frac14 \, \Phi'^2 \non\\ && +
\frac14 \, e^{-\Phi-2\, x}\, \left( e^{-2\, y}\, f'^2 + e^{2\, y}\, k'^2 + 2\, e^{2\, \Phi}\, F'^2 \right) \Big]
\eea
and the superpotential is given by
\beq
\label{superpotential}
W(\phi)=e^{4\, A - 2\, p-2\, x} + e^{4\, A+4\, p}\, \cosh y + \frac12\, e^{4\, A+4\, p-2\, x}\, \left( f \, (2\, P-F) + k\, F \right) \, .
\eeq

The background fields are given by the Klebanov--Strassler solution~\cite{Klebanov:2000hb}
 \bea \label{KSbackground}
 e^{x_0}&=& \frac14 \, h(\tau)^{1/2} \, \left( \tfrac12 \, \sinh(2\, \tau) - \tau \right)^{1/3} \, , \non \\
 e^{y_0}&=&\tanh(\tau/2) \, , \non \\
 e^{6\, p_0}&=&  24\, \frac{\left( \tfrac12 \, \sinh(2 \, \tau) - \tau \right)^{1/3}}{ h(\tau) \, \sinh^2\tau}  \, , \non \\
 e^{6\, A_0}&=&\frac{1}{3 \cdot 2^9} \, h(\tau)\, \left( \tfrac12 \, \sinh(2 \, \tau) - \tau \right)^{2/3}\, \sinh^2\tau \, , \\
 f_0&=& - P \, \frac{\left( \tau \, \coth \tau -1 \right)\, \left( \cosh \tau -1 \right)}{\sinh \tau} \, , \non\\
 k_0&=&- P\, \frac{\left(\tau \, \coth \tau -1 \right)\, \left(\cosh \tau +1 \right)}{\sinh \tau} \, , \non \\
 F_0&=& P\, \frac{\left(\sinh \tau -\tau \right)}{\sinh \tau} \, , \non \\
 \Phi_0&=&0 \, .\non 
 \eea
 The warp factor $h$ cannot be found analytically. Its expression as an integral is given below in equation~\eqref{hKS}.

\subsection{The first--order deformation: $\txi_a$ equations}

As in~\cite{Bena:2009xk} we shift to a slightly more convenient basis $\txi_a$, defined as
\be
\txi_a\equiv\left(3\xi_1-\xi_3+\xi_4,\xi_2,-3\xi_1+2\xi_3-\xi_4,-3\xi_1+\xi_3-2\xi_4,\xi_5+\xi_6,\xi_5-\xi_6,\xi_7,\xi_8 \right) \, .
\ee
The equations in the order in which we solve them, are\footnote{We have accounted for two misprints in the published version of~\cite{Bena:2009xk}.}
 \bea
 \txi_1'&=&e^{-2\, x_0} \, \left[ 2\, P\, f_0 - F_0\, \left( f_0 - k_0 \right) \right]\, \txi_1 \label{txi1eq} \\
 \txi_4'&=& -e^{-2\, x_0}\, \left[ 2\, P\, f_0 - F_0\, \left( f_0 - k_0 \right) \right]\, \txi_1  \label{txi4eq} \\
\txi_5'&=&-\frac13 \, P\, e^{-2\, x_0}\, \txi_1 \label{txi5eq} \\
\txi_6'&=&-\txi_7-\frac13 \, e^{-2\, x_0} \, \left( P - F_0 \right)\, \txi_1  \label{txi6eq} \\
\txi_7'&=&-\sinh(2\, y_0)\, \txi_5 - \cosh(2\, y_0)\, \txi_6 + \frac16 \, e^{-2\, x_0}\, \left( f_0-k_0 \right)\, \txi_1 \label{txi7eq} \\
\txi_8'&=& \left( P \, e^{2\, y_0} - \sinh(2\, y_0)\, F_0 \right)\, \txi_5 + \left( P\, e^{2\, y_0} - \cosh(2\, y_0)\, F_0 \right)\, \txi_6 +\frac12 \, \left( f_0 - k_0 \right)\, \txi_7 \label{txi8eq} \\
\txi_3'&=& 3\, e^{-2\, x_0 - 6\, p_0}\, \txi_3 + \left[ 5\, e^{-2\, x_0 - 6\, p_0} - e^{-2\, x_0}\,  \left( 2\, P\, f_0 - F_0\, \left( f_0-k_0 \right)\, \right) \right]\, \txi_1 \label{txi3eq} \\
\txi_2'&=& \txi_2\, \cosh y_0 + \frac13 \, \sinh y_0 \, \left( 2\, \txi_1 + \txi_3 + \txi_4 \right) \non \\
&& + 2\, \left[ \left( P\, e^{2\, y_0} - \cosh(2\, y_0)\, F_0 \right)\, \txi_5 + \left( P\, e^{2\, y_0} - \sinh(2\, y_0)\, F_0 \right)\, \txi_6 \right] \label{txi2eq}\, .
   \eea

The key development we present here is to solve for all the $\txi_a$ in terms of two simple integrals, one of which is the KS warp factor:
 \begin{align}\label{hKS}
h(\tau) &=  h_0 - 32\, P^2\, \int_0^{\tau} \frac{u\, \coth u - 1 }{\sinh^2 u}\, \left( \cosh u \sinh u-  u \right)^{1/3}\, du \, , \\
j(\tau)&
= \int^{\tau} \frac{du}{\left( \cosh u \, \sinh u - u \right)^{2/3}}  \ , \label{Gintegral}
 \end{align}
with $h_0=18.2373 \, P^2$ a numerical constant.

In solving the system of $\txi$ equations, we make the following key observations. In the equations for $\txi_1$ and $\txi_4$, we note that  
\be
e^{-2\, x_0}\, \slb 2\, P\, f_0 - F_0\, \left( f_0 - k_0 \right) \srb = \frac{h'}{h} \, .
\ee
This implies
\beq \label{txi1txi4}
\txi_1 =X_1 \, h(\tau) \, , \quad \ 
\txi_4= - X_1 \, h(\tau) + X_4 \ .
\eeq
To obtain $\txi_8$ we use the relations
\bea
P\, e^{2\, y_0} - \sinh (2\, y_0)\, F_0 &=& -\half \, \left( f_0 + k_0 \right)'  \, , \\
P\, e^{2\, y_0} - \cosh (2\, y_0)\, F_0 &=& -\half \, \left( f_0 - k_0 \right)'  \, ,
\eea
which yields
\bea
\txi_8' &=& -\half \left( f_0 + k_0 \right)' \, \txi_5 - \half \, \left( f_0 - k_0 \right)' \, \txi_6 + \frac12 \, \left( f_0 - k_0 \right)\, \txi_7 \, .
\eea
Integrating by parts and using (\ref{txi6eq}) and (\ref{txi1txi4}),
we get
\beq
\txi_8' = - \half \, \Blp \left( f_0 + k_0 \right) \, \txi_5 \Brp'  - \half \, \Blp \left( f_0 - k_0 \right) \, \txi_6 \Brp'  - \frac{1}{6}\, X_1 \, h' \, .
\eeq
This easily integrates to
\be\label{txi8}
\txi_8= -\half \, \left( f_0 + k_0 \right) \, \txi_5 - \half \, \left( f_0 - k_0 \right) \, \txi_6 - \frac{1}{6}\, X_1 \, h(\tau) + X_8 \, .
\ee
For $\txi_3$ we observe that 
\beq
e^{-2\, x_0 - 6\, p_0} &=& \frac{4}{3}\, \frac{\sinh^2 \tau}{\sinh 2\, \tau - 2 \, \tau} 
= \frac{1}{3}\, \frac{\txi'_{3,h} }{\txi_{3,h} } \, ,
\eeq
where $\txi_{3,h}$ is the solution to the homogeneous equation, namely 
\be
\txi_{3,h} = \sinh 2\, \tau - 2\, \tau \, .
\ee
As a result, 
\bea
\txi_3' &=& 3\, e^{-2\, x_0 - 6\, p_0}\, \txi_3 +  \left[ 5\, e^{-2\, x_0 - 6\, p_0} - e^{-2\, x_0}\, \left( 2\, P\, f_0 - F_0\, \left( f_0 - k_0 \right) \right) \right] \, \txi_1 \label{txi3eq} \non\\
&=& 3\, e^{-2\, x_0 - 6\, p_0}\, \txi_3 + X_1 \, \Blp \frac{5}{3} \, \frac{h\, \txi'_{3,h}}{\txi_{3,h}} - h' \Brp \, .
\eea
Henceforth,
\bea
\txi_3 =- \frac{5}{3}\, X_1 \, h(\tau) + \frac{2}{3}\, X_1 \, \txi_{3,h}\, \int^{\tau} du \, \frac{h'}{\txi_{3,h}}  + X_3\, \txi_{3,h} \, .
\eea
The above observations allow us to write the full solution in terms of two simple integrals. We collect here the full solution
\bea
\txi_1 &=& X_1 \, h(\tau) \, , \label{tX1}  \\
\txi_3 &= &\, - \frac{5}{3}\, X_1\, h(\tau) - \frac{32}{3}\, P^2\, X_1\, \text{csch}^2 \tau \, \left(\sinh \tau \, \cosh \tau - \tau \right)^{4/3} 
\non\\ 
&& - \frac{128}{9}\, P^2\, X_1\, \left( \sinh \tau \, \cosh \tau - \tau \right) \, j(\tau) + 2 X_3\, \left(\cosh \tau \sinh \tau - \, \tau \right) \, , \label{tX3} \\
\txi_4&=& - X_1 \, h(\tau) + X_4\, , \label{tX4} \\
\txi_5 &=& - \frac{16\, P}{3}\, X_1\, j(\tau) + X_5 \, ,\label{tX5} \\
\txi_6 &=& -\frac{1}{\sinh \tau}\, \lam_6(\tau) -\frac{\cosh \tau \sinh \tau -\, \tau}{2 \, \sinh \tau}\, \lam_7(\tau) \, , \label{txi6} \\
\txi_7 &=& -\frac{\cosh \tau}{\sinh^2 \tau}\, \lam_6(\tau) + \frac{-3+\cosh 2\, \tau + 2\, \tau\, \coth \tau}{4\, \sinh \tau}\, \lam_7(\tau)  \, , \label{txi7} \\
 \txi_8&=& P\, \left( \tau \, \coth \tau - 1 \right)\, \coth \tau \, \txi_5 - P\, \frac{\tau \, \coth \tau - 1}{\sinh \tau}\, \txi_6 - \frac{1}{6}\, X_1 \, h(\tau) + X_8 \, ,\label{tX8}
\eea
where
\bea
\lambda_6(\tau) &=&X_6 + \frac{1}{2}\, \left( - \tau + \coth \tau - \tau \, \coth^2 \tau \right)\, \txi_5(\tau) + \frac{1}{6}\, \frac{X_1}{P}\, h(\tau) \, , \\
\lambda_ 7(\tau) &=& X_7 - \csch^2\tau \, \txi_5(\tau) + \frac{16}{3}\, P\, X_1\, \text{csch}^2 \tau \, \left(\cosh \tau \, \sinh \tau - \tau \right)^{1/3} \non\\ && + \frac{64}{9}\, P\, X_1\, j(\tau) \,  .
\eea

Finally, $\txi_2$ can be obtained through the zero--energy condition (\ref{ZEC}) or just by direct integration of (\ref{txi2eq}). The latter will introduce another integration constant, $X_2$, that one could then determine as some combination of the other ones via the zero--energy condition. We find
\begin{align}
\txi_2 = & \, -\frac{2}{3}\, X_3\, \tau \, \cosh \tau + \frac{1}{3}\, X_4\, \cosh \tau + P\, X_6\, \text{csch}\tau \, \left( \coth \tau - \tau \, \text{csch}^2 \tau \right) \non\\ & + P\, X_5\, \text{csch} \tau \, \left( 1 - 2\, \tau \, \text{coth} \tau + \tau^2 \, \text{csch}^2 \tau \right) + X_2 \, \sinh \tau \non\\ & + \frac{1}{2}\, P\, X_7\, \left( - 2\, \tau \, \coth^3 \tau + \text{csch}^2 \tau + \tau^2\, \text{csch}^4 \tau\right)\, \sinh \tau \non\\ & - \frac{1}{108}\, X_1\, \bigg[ 3 \, \text{csch}^3 \tau \, h(\tau) \, \left( 6\, \tau - 5\, \sinh 2\, \tau + \sinh 4\, \tau \right) \non\\ & + 2\, P^2\, \text{csch}^5 \tau \, \big( - 15 + 24\, \tau^2 + 16\, \cosh 2\, \tau - \cosh 4\, \tau - 32\, \tau \, \sinh 2\, \tau + 4\, \tau \, \sinh 4\, \tau \big)\non\\ & \times \Big[ 4 \, \sinh^2 \tau \, j(\tau)  - 6\, \left( \cosh \tau \,  \sinh \tau - \tau \right)^{1/3} \Big] \bigg]
\end{align}

The zero--energy condition then amounts to\footnote{The relation between the $X_a$ integration constants in this paper and the ones in~\cite{Bena:2009xk} depends on the lower limit of the $j(\tau)$ integral (\ref{Gintegral}); it is not hard to check that this zero--energy condition is the same as that of~\cite{Bena:2009xk}.  }
\beq \label{ZECexplicit}
X_2 - \frac{2}{3}\, X_3 - P\, X_5 - \frac{3}{2}\, P\, X_7 = 0 \, .
\eeq

\subsection{The first--order deformation: $\tphi^{a}$ equations}

The analytic expressions we obtain for the eight $\tphi^a$ modes are all double integrals, except for $\tphi^4$ where we obtain a triple integral. This is a considerable improvement over
all previous works. Depending on the reader's taste, the expressions may appear somewhat cumbersome but we find them crucial for explicit numerical computations 
which will appear in a companion paper~\cite{UStoappear}.

To solve the system of $\phi^a$, we also use a shifted basis~\cite{Bena:2009xk}
 \be
 \tphi_a=(x-2p-5A,y,x+3p, x-2p-2A ,f,k,F,\Phi) \ . \label{tphidef}
 \ee
The system of equations for the $\tphi^a$ modes is (in the order in which we actually solve them)
 \bea
  \tphi_8^{\prime} &=& - 4\, e^{-4\, A_0 - 4\, p_0}\, \txi_8 \label{phi8peq} \, , \\
  \tphi_2^{\prime} &=& - \cosh y_0\, \tphi_2 - 2\, e^{-4 \, A_0 - 4\, p_0}\, \txi_2 \label{phi2peq} \, , \\
 \tphi_3^{\prime} &=&  - 3\, e^{-6\,p_0-2\,x_0}\, \tphi_3 - \sinh y_0\, \tphi_2 - \frac16 \, e^{-4\, A_0 - 4\, p_0}\,  \left( 9\, \txi_1 + 5\, \txi_3 + 2\, \txi_4 \right) \, , \label{phi3peq}\\
  \tphi_1^{\prime} &=&  2\, e^{-6\, p_0-2\, x_0}\, \tphi_3 - \sinh y_0 \, \tphi_2 + \frac{1}{6} \, e^{-4 \, A_0 - 4\, p_0}\, \left( \txi_1 + 3\, \txi_4 \right)  \, , \label{phi1peq}\\
 \tphi_5^{\prime} &=&   e^{2\,y_0}\, \left( F_0 - 2\, P \right)\, \left( 2\, \tphi_2 + \tphi_8 \right) + e^{2\, y_0}\, \tphi_7
 - 2\, e^{-4 \, A_0 - 4\, p_0 + 2\, x_0 + 2\, y_0}\, \left( \txi_5 + \txi_6 \right) \, , \label{phi5peq}\\
 \tphi_6^{\prime}  &=&  e^{-2\,y_0}\, \left[ F_0 \, \left( 2 \, \tphi_2 - \tphi_8 \right) - \tphi_7 \right] - 2\, e^{-4\, A_0 - 4\,p_0 + 2\, x_0 - 2\, y_0}\, \left( \txi_5 - \txi_6 \right) \, , \label{phi6peq}\\
 \tphi_7^{\prime} &=& \half \, \Blp \tphi_5 - \tphi_6 + \left( k_0 - f_0 \right)\, \tphi_8 \Brp - 2 \, e^{-4\,A_0 -4\,p_0 + 2\, x_0}\, \txi_7 \, , \label{phi7peq} \\
  \tphi_4^{\prime} &=& \frac{1}{5}\, e^{-2\, x_0} \left[ f_0\, \left( 2\, P - F_0 \right) + k_0\, F_0 \right]\, \left( 2\, \tphi_1 - 2\, \tphi_3 - 5\, \tphi_4 \right) + \frac{1}{2}\, e^{-2\,x_0}\, \left( 2\, P - F_0 \right)\, \tphi_5 \non  \\
 && + \frac{1}{2}\, e^{-2\,x_0}\, F_0\, \tphi_6 + \frac{1}{2}\, e^{-2\,x_0}\, \left( k_0 - f_0 \right)\, \tphi_7 - \frac{1}{3}\, e^{- 4 \,A_0 - 4\, p_0}\, \txi_1 \, . \label{phi4peq}
 \eea

\subsubsection{The $\tphi^{8}$ solution}
By directly integrating~\eqref{phi8peq} and a little bit of massaging, we arrive at
\begin{align}\label{phi8}
\tphi_8 =&\, Y_8 - 64\, X_8\, j(\tau) + \frac{X_7}{P}\, h(\tau) \non\\ & - 64\, P\, X_6\, \int^{\tau} \frac{\left( u\, \coth u - 1 \right)}{\sinh^2 u \, \left( \cosh u \, \sinh u - u \right)^{2/3}} \, d u \non\\ & + \frac{2}{P}\, h(\tau)\, \txi_5(\tau) + \frac{16}{3}\, X_1\, \text{csch}^2 \tau \left( \cosh \tau \, \sinh \tau - \tau \right)^{1/3}\, h(\tau) \non\\ & + \frac{64}{9}\, X_1\, h(\tau) \, j(\tau) 
+ \frac{64}{3}\, X_1\, \int^{\tau} \frac{\left( \sinh^2 u + 1 - u \, \coth u \right)}{\sinh^2 u \, \left( \cosh u \, \sinh u - u \right)^{2/3}}\, h(u)\, d u \, .
\end{align}

\subsubsection{The $\tphi^{2}$ solution}
The solution to equation~\eqref{phi2peq} is given by 
\be\label{phi2}
\tphi_2 = \text{csch} \tau \, \Lambda_2(\tau) \, ,
\ee
where 
\begin{align}\label{Lambda2}
\Lambda_2(\tau) = & \, Y_2 - 16\, P\, X_7\, \int^{\tau} \frac{\left(- 2\, u \, \coth^3 u + \text{csch}^2 u + u^2\, \text{csch}^4 u \right)\, \sinh^2 u}{\left(\cosh u \, \sinh u - u \right)^{2/3}} \, d u \non\\ & - 32\, P\, X_6\, \int^{\tau} \frac{\coth u - u \, \text{csch}^2 u}{\left( \cosh u \, \sinh u - u \right)^{2/3}} \, d u - 32\, P\, X_5\, \int^{\tau} \frac{1- 2\, u\, \coth u + u^2\, \text{csch}^2 u}{\left( \cosh u \, \sinh u - u \right)^{2/3}} \, d u \non\\ & - \frac{32}{3}\, X_4\, \int^{\tau} \frac{\cosh u \, \sinh u}{\left( \cosh u \, \sinh u - u \right)^{2/3}} \, d u + \frac{64}{3}\, X_3\, \int^{\tau} \frac{u \, \cosh u \, \sinh u}{\left( \cosh u \, \sinh u - u \right)^{2/3}} \, d u \non\\ & - 48\, X_2\, \left( \cosh \tau \, \sinh \tau - \tau \right)^{1/3} + \frac{8}{9}\, X_1\, \int^{\tau} \frac{6\, u - 5\, \sinh 2\, u + \sinh 4\, u}{\sinh^2 u \, \left(\cosh u \, \sinh u - u \right)^{2/3}}\, h(u)\, d u \non\\ & 
- \frac{32}{9}\, P^2\, X_1\, \int^{\tau} \frac{- 15 + 24\, u^2 + 16\, \cosh 2\, u - \cosh 4\, u - 32\, u \, \sinh 2\, u + 4\, u \, \sinh 4\, u }{\sinh^4 u \, \left( \cosh u \, \sinh u - u \right)^{1/3}} \, d u \non\\ & 
+ \frac{64}{27}\, P^2\, X_1\, \int^{\tau} \frac{-15 + 24\, u^2 + 16\, \cosh 2\, u - \cosh 4\, u - 32\, u \, \sinh 2\, u + 4\, u \, \sinh 4\, u}{\sinh^2 u \, \left( \cosh u \, \sinh u - u \right)^{2/3}}\, j(u)\,  d u \, .
\end{align}
%
\subsubsection{The $\tphi^{3}$ solution}
Equation~\eqref{phi3peq} is solved by
\be\label{phi3}
\tphi_3(\tau) = \frac{1}{\sinh 2\, \tau - 2\, \tau}\, \Lambda_3 (\tau) \, ,
\ee
where $\Lambda_3$ is specified as
\begin{align}
\Lambda_3 =&\, Y_3\, -\frac{32}{3}\, X_4\, \int^{\tau} \left( \cosh u \, \sinh u - u \right)^{1/3} \, du - \frac{112}{3}\, X_1\, \int^{\tau} \left( \cosh u \, \sinh u - u \right)^{1/3}\, h(u)\, d u \non\\ & - \frac{80}{3}\, \int^{\tau} \left( \cosh u \, \sinh u - u \right)^{1/3}\, \txi_3(u) \, d u + 2\, \tau \, \coth \tau \, \Lambda_2(\tau) - 2\, \int^{\tau} u\, \coth u \, \Lambda_2^{\prime}(u)\, du \, .
\end{align}
Expanding and simplifying this expression, it becomes
\begin{align}\label{Lambda3}
\Lambda_3 = &\, Y_3 + 32\, P\, X_7\, \int^{\tau} \frac{u \, \cosh u \, \left( - 2 \, u \, \coth^3 u + \text{csch}^2 u + u^2\, \text{csch}^4 u \right) \, \sinh u}{\left( \cosh u \, \sinh u - u \right)^{2/3}} \, d u \non\\ & + 64\, P\, X_6\, \int^{\tau} \frac{u\, \coth u \, \left( \coth u - u \, \text{csch}^2 u \right)}{\left( \cosh u \, \sinh u - u \right)^{2/3}} \, d u \non\\ & + 64\, P\, X_5\, \int^{\tau} \frac{ u \, \coth u \, \left(1 - 2\, u \, \coth u + u^2 \, \text{csch}^2 u \right)}{\left( \cosh u \, \sinh u - u \right)^{2/3}} \, d u \non\\ & + \frac{32}{3}\, X_4\, \left\{ 2 \, \int^{\tau} \frac{u \, \cosh^2 u}{\left(\cosh u \, \sinh u - u \right)^{2/3}} \, d u - \int^{\tau} \left( \cosh u \, \sinh u - u \right)^{1/3} \, d u \right\} \non\\ & - \frac{32}{3}\, X_3 \, \left\{ 5\, \int^{\tau} \left( \cosh u \, \sinh u - u \right)^{4/3} \, d u + 4\, \int^{\tau} \frac{u^2\, \cosh^2 u}{\left( \cosh u \, \sinh u - u \right)^{2/3}} \, d u \right\} \non\\ & + 2\, \tau \, \coth \tau \, \Lambda_2(\tau) + 64\, X_2\, \int^{\tau} \frac{u \, \cosh u \, \sinh u}{\left( \cosh u \, \sinh u - u \right)^{2/3}} \, d u \non\\ & + \frac{64}{9}\, X_1\, \int^{\tau} \left( \cosh u \, \sinh u - u \right)^{1/3} h(u)\, d u \non\\ &
+\frac{10240}{27}\, P^2\, X_1\, \int^{\tau} \left( \cosh u \, \sinh u - u \right)^{4/3}\, j(u)\, d u \non\\ & + \frac{2560}{9}\, P^2\, X_1\, \int^{\tau} \text{csch}^2 u \, \left( \cosh u \, \sinh u - u \right)^{5/3}\, d u \non\\ & - \frac{16}{9}\, X_1\, \int^{\tau} \frac{u \, \coth u \, \text{csch}^2 u \, \left( 6\, u - 5\, \sinh 2\, u + \sinh 4\, u \right)}{\left( \cosh u \, \sinh u - u \right)^{2/3}} \, h(u)\, d u \non\\ &
+ \frac{64}{9}\, P^2\, X_1\, \int^{\tau} \frac{u \, \coth u \, \left( - 15 + 24\, u^2 + 16\, \cosh 2\, u -\cosh 4\, u - 32\, u \, \sinh 2\, u + 4\, u \, \sinh 4\, u \right)}{\sinh^4 u \, \left(\cosh u \, \sinh u - u \right)^{1/3}} \, d u \non\\ &
-\frac{128}{27}\, P^2\, X_1\, \int^{\tau} \left( - 15 + 24\, u^2 + 16\, \cosh 2\, u - \cosh 4\, u - 32\, u \, \sinh 2\, u + 4\, u \, \sinh 4\, u \right)\, \non\\ & \times \frac{u \, \coth u \, \text{csch}^2 u}{\left( \cosh u \, \sinh u - u \right)^{2/3}}\, j(u) \, d u \, .
\end{align}

\subsubsection{The $\tphi^{1}$ solution}

Next comes $\tphi_1$ which we express concisely in terms of $\Lambda_2$ and $\tphi_3$:
\begin{align}\label{phi1}
\tphi_1 = &\, Y_1 + \frac{40}{9}\, X_4\, j(\tau) - \frac{2}{3}\,
\tphi_3(\tau) - \frac{160}{9}\, X_3 \, \int^{\tau} \left( \cosh u \,
  \sinh u - u \right)^{1/3} \, d u \non\\ & + \frac{5}{3}\, \int \coth
u \, \Lambda_2^{\prime}(u)\, d u  -\frac{5}{3}\coth \tau \, \Lambda_2(\tau)  + \frac{2560}{27}\, P^2\, X_1\, \int^{\tau} \text{csch}^2 u \, \left( \cosh u \, \sinh u - u \right)^{2/3}\, d u \non\\ & + \frac{10240}{81}\, P^2\, X_1\, \int^{\tau} \left( \cosh u \, \sinh u - u \right)^{1/3} \, j(u)\, d u  - \frac{80}{27}\, X_1\, \int^{\tau} \frac{h(u)}{\left(\cosh u \, \sinh u - u \right)^{2/3}} \, d u \, .
\end{align}
\subsubsection{The $(\tphi^{5},\tphi^{6},\tphi^{7})$ solutions}

The fields $\tphi_{5,6,7}$ are determined by a system of coupled ordinary differential equations. The homogeneous solutions are easily found and then we apply the Lagrange method of variation of parameters to find the following expressions
\bea
\tphi_5 &=& \frac{1}{2}\, \text{sech}^2(\tau/2)\, \left[ \tau + 2\, \tau \, \cosh \tau - \left( 2 + \cosh \tau \right)\, \sinh \tau \right]\, \Lambda_5(\tau) + \frac{1}{1+\cosh \tau}\, \Lambda_6(\tau) + \Lambda_7(\tau) \, ,\non\\ \label{phi5} \\
\tphi_6 &=& \left[ \tau \, \left( 2 - \frac{1}{1 - \cosh \tau} \right) - \coth(\tau/2) + \sinh \tau \right]\, \Lambda_5(\tau) + \frac{1}{1-\cosh \tau}\, \Lambda_6(\tau) + \Lambda_7(\tau) \, , \label{phi6}\\
\tphi_7 &=& \left( - \cosh \tau + \tau \, \text{csch} \tau \right)\, \Lambda_5(\tau) - \text{csch}\tau \, \Lambda_6(\tau) \, , \label{phi7}
\eea
where
\begin{align}\label{Lambda5}
\Lambda_5 = &\, Y_5 - \frac{1}{2}\, P\, \left(\tau \, \coth \tau -1 \right) \, \text{csch}^2 \tau \, \tphi_8(\tau) - 32\, P\, \int^{\tau} \frac{\left( u \, \coth u - 1\right)\, \text{csch}^2 u}{\left( \cosh u \, \sinh u - u \right)^{2/3}} \, \txi_8(u)\, d u \non\\ & +\frac{1}{4}\, X_7\, \int^{\tau} \text{csch}^4 u \, \left[ 2\, u \, \left( 2 + \cosh 2\, u \right) - 3\, \sinh 2\, u \right]\, h(u)\, d u -X_6\, \int^{\tau} \frac{2 + \cosh 2\, u}{\sinh^4 u}\, h(u)\, d u \non\\ & + \int^{\tau} \text{csch}^2 u \, \left[ - 3\, \coth u + u \, \left( 2 + 3\, \text{csch}^2 u \right) \right]\, h(u)\, \txi_5(u)\, d u - \frac{1}{2}\, P\, \frac{ \cosh \tau \,  \sinh \tau - \tau }{\sinh^4 \tau}\, \Lambda_2(\tau) \non\\ & + \frac{1}{2}\, P\, \int^{\tau} \text{csch}^4 u \, \left( \cosh u \, \sinh u - u \right)\, \Lambda_2^{\prime}(u)\, d u - \frac{X_1}{6\, P}\, \int^{\tau} \left( 2 + \cosh 2\, u \right)\, \text{csch}^4 u \, h^2(u)\, d u \non\\ & + \frac{16}{9}\, P\, X_1\, \int^{\tau} \text{csch}^4 u \, \left[ 2\, u \, \left( 2 + \cosh 2\, u \right) - 3\, \sinh 2\, u \right]\, j(u)\, h(u)\, d u \non\\ & + \frac{4}{3}\, P\, X_1\, \int^{\tau} \text{csch}^6 u \, \left( \cosh u \, \sinh u - u \right)^{1/3}\, \left[ 2\, u\, \left( 2 + \cosh 2\, u \right) - 3\, \sinh 2\, u \right]\, h(u)\, d u  \, ,  
\end{align}
\begin{align}\label{Lambda6}
\Lambda_6 = &\, Y_6 - \frac{1}{2}\, P\, \left[ -\tau  + \coth \tau  +  \tau  \, \left( - 2 +  \tau  \, \coth  \tau  \right) \, \text{csch}^2  \tau  \right]\, \tphi_8( \tau ) 
\non\\&
- 32\, P\, \int^{\tau} \frac{\left[-  u  + \coth  u  +  u  \, \left( - 2 +  u  \, \coth  u  \right)\, \text{csch}^2  u  \right]}{\left(\cosh  u  \, \sinh  u  -  u  \right)^{2/3}}\, \txi_8( u )\, d u  
\non\\& 
+ \frac{1}{2}\, X_7\, \int^{\tau} \left[ \cosh 2\,  u  +\text{csch}^2  u  \, \left( 3 + 2\,  u ^2 - 6\,  u  \, \coth  u  + 3\,  u ^2\, \text{csch}^2  u  \right) \right]\, h( u ) \, d u  
\non\\&
+X_6\, \int^{\tau} \text{csch}^2  u  \, \left[ 3\, \coth  u  -  u  \, \left( 2 + 3\, \text{csch}^2  u  \right) \right]\, h( u )\, d u  
\non\\&
+\int^{\tau} \left[ 1 + \left( 3 + 2\,  u ^2 - 6\,  u  \, \coth  u  \right)\, \text{csch}^2  u  + 3\,  u ^2\, \text{csch}^4  u  \right]\, h( u )\, \txi_5( u )\, d u  
\non\\&
-\frac{1}{2}\, P\, \left[ 2\, \coth^2 \tau  \, \left( - 1 +  \tau  \, \coth  \tau  \right) + \text{csch}^2  \tau  -  \tau ^2\, \text{csch}^4  \tau  \right] \Lambda_2( \tau ) 
\non\\&
+ \frac{1}{2}\, P\, \int^{\tau} \left[ 2\, \coth^2  u  \, \left( - 1 +  u  \, \coth  u  \right) + \text{csch}^2  u  -  u ^2\, \text{csch}^4  u  \right]\, \Lambda_2^{\prime}( u )\, d u  
\non\\&
+ X_1\, \int^{\tau} \bigg\{ \frac{\text{csch}^4 u\, \left[ - 2\,  u  \left( 2 + \cosh 2\,  u  \right) + 3\, \sinh 2\,  u  \right]}{12\, P} h( u )\, + \frac{1}{36}\, P\, \text{csch}^6 u
\non\\&
\times \left[ 8 \, j(u)\, \sinh^2  u  + 6\, \left( \cosh  u  \, \sinh  u  -  u  \right)^{1/3} \right] \Big[ - 28 + 32\,  u ^2 + \left( 31 + 16\,  u ^2 \right)\, \cosh 2\,  u  \non\\ & - 4\, \cosh 4\,  u  + \cosh 6\,  u  - 48\,  u  \, \sinh 2\,  u  \Big] \bigg\} \, h(u)\, d u 
\end{align}
and
\begin{align}\label{Lambda7}
\Lambda_7 = &\, Y_7 + P\, \left[ -  \tau + \coth  \tau +  \tau \, \left( - 2 +  \tau \, \coth  \tau \right)\, \text{csch}^2  \tau \right] \, \tphi_8( \tau ) \non\\ & + 64\, P\, \int^{\tau} \frac{\left[ -  u + \coth  u +  u \, \left( - 2 +  u \, \coth  u \right)\, \text{csch}^2  u \right]}{\left( \cosh  u \, \sinh  u -  u \right)^{2/3} } \, \txi_8( u )\, d u \non\\ & + X_7\, \int^{\tau} \left[ - 1 + \left( - 3 - 2\,  u ^2 + 6\,  u \, \coth  u \right)\, \text{csch}^2  u - 3\,  u ^2\, \text{csch}^4  u \right]\, h( u ) \, d u \non\\ & + X_6\, \int^{\tau} \text{csch}^4  u \, \left[ 2\,  u \, \left( 2 + \cosh 2\,  u \right) - 3 \, \sinh 2\,  u \right] \, h( u )\, d u \non\\ & 
+ \int^{\tau} \left[ - 2 - 2\, \text{csch}^2  u \, \left( 3 + 2\,  u ^2 - 6\,  u \, \coth  u + 3\,  u ^2\, \text{csch}^2  u \right) \right]\, h( u )\, \txi_5( u )\, d u \non\\ &
- P\, \text{csch}^2 \tau \, \left( 1 - 2\,  \tau \, \coth  \tau +  \tau ^2\, \text{csch}^2  \tau \right)\, \Lambda_2( \tau ) \non\\ &
+ P\, \int^{\tau} \text{csch}^2 u \, \left( 1 - 2\,  u \, \coth  u +  u ^2\, \text{csch}^2  u \right)\, \Lambda_2^{\prime}( u )\, d u \non\\ & 
+ X_1\, \int^{\tau} \bigg\{ \frac{\text{csch}^4 u \, \left[ 2\,  u \, \left( 2 + \cosh 2\,  u \right) - 3\, \sinh 2\,  u \right]}{6\, P}\, h(u) - \frac{1}{9}\, P\, \text{csch}^6  u \, \non\\ & 
\times \left[ 8\, j(u) \,\sinh^2  u + 6\, \left( \cosh  u \, \sinh  u -  u \right)^{1/3} \right]\, \non\\ & 
\times \left[ - 9 + 16\,  u ^2 + 8\, \left( 1 +  u ^2 \right)\, \cosh 2\,  u + \cosh 4\,  u - 24\,  u \, \sinh 2\,  u \right] \bigg\} \, h(u)\, d u \, .
\end{align}
%

\subsubsection{The $\tphi^{4}$ solution}

While all the $\tphi^a$ modes so far have been double integrals,
we obtain for $\tphi_4$ a triple integral expression
\begin{align}
\tphi_4(\tau) = &\, \frac{1}{h(\tau)}\, \Big\{Y_4 - \frac{16}{3}\, X_1\, \int^{\tau} \frac{h(u)^2}{\left( \cosh u \, \sinh u - u \right)^{2/3}} \, d u + 32\, P\, \int^{\tau} \frac{\left( u\, \coth u - 1 \right)\, \text{csch}^2 u \, \Lambda_6(u)}{\left( \cosh u \sinh u - u \right)^{2/3}} \, d u \non\\ & + 16\, P\, \int^{\tau} \frac{\Lambda_7(u)}{\left(\cosh u \, \sinh u - u \right)^{2/3}} \, d u + \frac{32}{5}\, P\, \int^{\tau} \left( u\, \coth u - 1 \right)\, \text{csch}^2 u \, \left( \cosh u \, \sinh u - u \right)^{1/3}\, \non\\ & \times \left[ 5\, \Lambda_5(u) + 2\, P\, \left( - \tphi_1(u) + \tphi_3(u) \right) \right]\, du  \Big\} \, .
\end{align}
It may be possible that this expression can also be reduced to double integrals, but we could not find any obvious way to do it. This completes the solution to the system.

\section{Boundary Conditions and Anti--D3 Branes}
The deformation space  we have solved for is a fourteen--dimensional linear space\footnote{One of the $X$'s can be eliminated through the zero--energy condition~\eqref{ZECexplicit}, while $Y_1$ is a gauge degree of freedom corresponding to a constant rescaling of the 4D metric.} and has numerous solutions, out of which one has to fish out the possible solution for backreacted anti--D3 branes by imposing appropriate boundary conditions. 

The strategy for doing this was explained in detail in~\cite{Bena:2009xk}: one should eliminate all integration constants that give divergent fields in the IR or non-normalizable modes in the UV. Furthermore, to argue that the solution corresponds to D3 branes one should set the divergence in the warp factor perturbation (given by $\tphi_4$) to be commensurate with the divergence coming from the five--form. This should fix all the integration constants in terms of the would--be anti--D3 charge. Nevertheless, to obtain the precise values of these constants, one needs to relate the infrared and ultraviolet expansion parameters of the modes presented in this paper, and this can only be done by numerical integration.

In particular, this calculation will fix the constant $X_1$ in terms of the additional anti--D3 brane charge $N_{ {\overline{\rm D3}}}$ which is given by the UV value of the mode $\tphi^5+\tphi^6$.  Having done that, one can test whether the candidate solution describes indeed anti--D3 branes by using the universal form of the force exerted on a probe D3 brane by any solution to the equations~\cite{Bena:2009xk,Bena:2010ze}:
\be
F_{\rm D3}=\frac{32}{3}\, \frac{ X_1[N_{ {\overline{ \rm D3}}} ]}{\left( \sinh \, \tau \cosh \, \tau-  \, \tau \right)^{2/3}}
\ee
and comparing this with the force anti--D3 branes should exert \cite{Kachru:2003sx,Bena:2010ze}. The expressions for the $(\txi_a,\tphi^a)$ obtained here in terms of single and double integrals turn out to be crucial in this endeavor, on which we intend to report soon.

\vspace{.2cm}

\noindent {\bf Acknowledgements}:
\noindent The work of G.~G.~and S.~M.~is supported by a Contrat de Formation par la Recherche of CEA/Saclay. The work of I.~B., M.~G.~and N.~H.~is supported by the DSM CEA/Saclay, the ANR grants 07--CEXC--006 and 08--JCJC--0001--0, and by the ERC Starting Independent Researcher Grants 240210 -- String--QCD--BH and 259133 -- ObservableString.

\providecommand{\href}[2]{#2}\begingroup\raggedright\endgroup

\end{document}